\documentstyle [12pt] {article}

\catcode`@=12
\begin{document}
\thispagestyle{empty}
\begin{flushright} UCRHEP-T253\\May 1999\
\end{flushright}
\vskip 0.5in
\begin{center}
{\Large \bf Spontaneous Supersymmetric Generation\\
of an Indeterminate Mass Scale and\\
a Possible Light Sterile Neutrino\\}
\vskip 1.5in
{\bf Ernest Ma\\}
\vskip 0.1in
{\sl Physics Department, Univeristy of California,\\}
{\sl Riverside, CA 92521, USA\\}
\end{center}
\vskip 1.5in
\begin{abstract}\
If a global continuous symmetry of a supersymmetric field theory is 
spontaneously broken while preserving the supersymmetry, the resulting 
theory has a massless superfield.  One of its two bosonic degrees of 
freedom is the familiar phase rotation of the usual massless Nambu-Goldstone 
boson, but the other is a scale transformation.  An indeterminate mass 
scale is thus generated.  In the fermion sector, a seesaw texture appears 
which may be suitable for a possible light sterile neutrino.  This feature 
persists even after the gauging of the continuous symmetry or the breaking 
of the supersymmetry to resolve the aforementioned mass-scale ambiguity.
\end{abstract}
\newpage
\baselineskip 24pt

The physical mass scales of a renormalizable quantum field theory are 
expected to be determined by its explicit parameters, such as in quantum 
electrodynamics, or by the structure of its interactions, such as in quantum 
chromodynamics.  In either case, even if the vacuum has nontrivial topology, 
the physical mass scales of the theory are uniquely determined.  Consider 
the textbook example of the U(1) scalar model, with the potential
\begin{equation}
V_1 = m^2 \phi^* \phi + {1 \over 2} \lambda (\phi^* \phi)^2.
\end{equation}
If $m^2 < 0$, then the global U(1) symmetry is broken by the vacuum 
expectation value (VEV) of the scalar field $\phi$, {\it i.e.}
\begin{equation}
\langle \phi \rangle \equiv v = \sqrt {-m^2 \over \lambda} ~e^{i \theta}.
\end{equation}
Redefining
\begin{equation}
\phi \equiv v + e^{i \theta} \left( {H + i \xi \over \sqrt 2} \right),
\end{equation}
the physical theory is then given by
\begin{equation}
V_2 = {1 \over 2} m_H^2 H^2 + {1 \over 2} m_H \sqrt \lambda H (H^2 + \xi^2) + 
{1 \over 8} \lambda (H^2 + \xi^2)^2,
\end{equation}
where $m_H^2 = -2m^2$, {\it i.e.} the mass scale of this theory is still 
determined by the input $m^2$.  As is well-known, the massless Nambu-Goldstone 
boson $\xi$ is a manifestation \cite{1} of the rotational degree of 
freedom $e^{i\theta}$ of Eq.~(2).

Consider now a supersymmetric U(1) model.  Assume the spontaneous breakdown 
of the global U(1) symmetry, but not the supersymmetry \cite{2}.  The 
Goldstone theorem \cite{3} requires the existence of a massless 
Nambu-Goldstone boson, which must then be part of a massless superfield, 
{\it i.e.} there must exist a massless complex scalar field together with 
its massless fermionic superpartner.  One of its two bosonic degrees of 
freedom is the analog of $e^{i\theta}$, but the other is a scale 
transformation $e^\alpha$, where $\alpha$ is a real parameter.  It is 
clear that a model with just one VEV cannot have this property, because 
its ground state is obviously not invariant under the transformation 
$v \to v e^\alpha$.  On the other hand, if there are two VEVs, it may 
become possible to have $v_1 \to v_1 e^\alpha$ and $v_2 \to v_2 e^{-\alpha}$ 
so that their product $v_1 v_2$ remains unchanged.  As shown below, that is 
exactly what happens in the simplest realization of this phenomenon.  Since 
$|v_1|^2 + |v_2|^2$ is unconstrained by the explicit parameters of such a 
model, an indeterminate mass scale is generated.

The simplest model which exhibits the behavior under consideration has three 
superfields: $\phi_1$ and $\phi_2$ transform oppositely under U(1) and 
$\chi$ is trivial under it.  The most general superpotential is given by
\begin{equation}
W = \mu \phi_1 \phi_2 + f \phi_1 \phi_2 \chi + {1 \over 2} m \chi^2 + 
{1 \over 3} h \chi^3,
\end{equation}
from which the following scalar potential is obtained:
\begin{equation}
V = |\mu + f \chi|^2 \left( |\phi_1|^2 + |\phi_2|^2 \right) + |f \phi_1 \phi_2 
+ m \chi + h \chi^2|^2.
\end{equation}
There are three $V = 0$ solutions: (1) $\phi_1 = \phi_2 = \chi = 0$; 
(2) $\phi_1 = \phi_2 = 0$, $\chi = -m/h$; and (3) $\chi = -\mu/f$, 
$\phi_1 \phi_2 = m \mu/f^2 - h \mu^2/f^3$.  Whereas U(1) is unbroken in 
the first two cases, it is spontaneously broken in the third, resulting in 
the existence of ``flat directions'' \cite{4}.

Let $\phi_{1,2}$ and $\chi$ be shifted to $v_{1,2} + \phi_{1,2}$ and 
$u + \chi$ in Eq.~(6), where
\begin{equation}
u = -{\mu \over f}
\end{equation}
and
\begin{equation}
v_1 v_2 = {m \mu \over f^2} - {h \mu^2 \over f^3},
\end{equation}
then $V$ yields the following mass terms:
\begin{eqnarray}
V_2 &=& \left[ |f|^2 \left( |v_1|^2 + |v_2|^2 \right) + |m + 2 h u|^2 \right] 
|\chi|^2 + |f|^2 |v_2 \phi_1 + v_1 \phi_2|^2 \nonumber \\ &+& \left[ f^* 
(m + 2 h u)(v_2^* \phi_1^* + v_1^* \phi_2^*) \chi + h.c. \right].
\end{eqnarray}
This shows clearly that the linear combination
\begin{equation}
\zeta \equiv {-v_1^* \phi_1 + v_2^* \phi_2 \over \sqrt {|v_1|^2 + |v_2|^2}}
\end{equation}
is massless, whereas
\begin{equation}
\eta \equiv {v_2 \phi_1 + v_1 \phi_2 \over \sqrt {|v_1|^2 + |v_2|^2}}
\end{equation}
and $\chi$ are massive with their mass-squared matrix given by
\begin{equation}
{\cal M}^2 = \left[ \begin{array} {c@{\quad}c} |A|^2 & A^* B \\ A B^* & 
|A|^2 + |B|^2 \end{array} \right],
\end{equation}
where
\begin{equation}
A = f \sqrt {|v_1|^2 + |v_2|^2}, ~~~ B = m + 2 h u.
\end{equation}
The constraint of Eq.~(8) applies only to the product $v_1 v_2$, hence 
$A$ of Eq.~(13) is an indeterminate parameter.

Unlike the usual Nambu-Goldstone boson, $\zeta$ of Eq.~(10) is a \underline 
{complex} scalar field.  One of its two degrees of freedom corresponds to 
having
\begin{equation}
v_1 \to v_1 e^{i\theta}, ~~~ v_2 \to v_2 e^{-i\theta}
\end{equation}
in analogy to the familiar invariance of the ground state with respect 
to a phase rotation, but the other is a scale transformation, {\it i.e.}
\begin{equation}
v_1 \to v_1 e^\alpha, ~~~ v_2 \to v_2 e^{-\alpha}.
\end{equation}
Hence the individual values of $|v_1|$ and $|v_2|$ are not separately 
determined.  This is the consequence of the spontaneous breakdown of a 
global continuous symmetry together with the assumed preservation of the 
supersymmetry.  It is also easily shown that the fermion partners of 
$\eta$ and $\chi$ have the mass matrix
\begin{equation}
{\cal M} = \left[ \begin{array} {c@{\quad}c} 0 & A \\ A & B \end{array} 
\right],
\end{equation}
hence ${\cal M}^\dagger {\cal M} = {\cal M}^2$ of Eq.~(12) as expected.

The superpotential $W$ of Eq.~(5) has 4 parameters: $f$, $h$, $\mu$, and 
$m$.  The spontaneously broken theory has instead 5 parameters: 
$f$, $h$, $u$, $v_1$, and $v_2$.  Whereas $u$ and $v_1 v_2$ are constrained 
by Eqs.~(7) and (8), $|v_1|^2 + |v_2|^2$ is not.  The new superpotential is 
then given by
\begin{eqnarray}
W' &=& f \sqrt {|v_1|^2 + |v_2|^2} \eta \chi + {1 \over 2} (m + 2 h u) \chi^2 
+ {1 \over 3} h \chi^3 \nonumber \\ &+& {f \chi \over |v_1|^2 + |v_2|^2} 
\left[ v_1^* v_2^* \eta^2 + (|v_2|^2 - |v_1|^2) \eta \zeta - v_1 v_2 \zeta^2 
\right].
\end{eqnarray}
If it is not known that $W$ is the antecedent of $W'$, one may worry that 
the massless superfield $\zeta$ would not stay massless in the presence 
of interactions.  As it is, because of the Goldstone theorem, all such 
higher-order effects do in fact cancel and $\zeta$ is indeed massless.  
I have checked this explicitly to one-loop order, and have ascertained 
that the cancellation works for arbitrary values of $|v_1|^2 + |v_2|^2$. 
I note also that this phenomenon occurs in general whenever a global 
continuous symmetry is reduced in rank by one spontaneously while 
preserving the supersymmetry.

Since $|v_1|^2 + |v_2|^2$ is not fixed by the input parameters $f$, $h$, 
$\mu$, and $m$ of $W$, the scale of the spontaneous breakdown of the global 
U(1) symmetry is arbitrary, subject only to the algebraic inequality 
$|v_1|^2 + |v_2|^2 \geq 2|v_1 v_2|$.  This is an unusual phenomenon which 
may be relevant to cosmology if supersymmetry is a good description of 
fundamental interactions in the early Universe.  It allows for the 
possibility of different domains, not just of phase, but of scale.  On 
the other hand, this ambiguity of scale may be just a curiosity and is 
naturally eliminated in a realistic theory.  There are two ways, as 
discussed below.

One way is to promote the global U(1) symmetry to a local U(1) symmetry 
\cite{5}, so that there are now gauge interactions which generate an extra 
term in the scalar potential:
\begin{equation}
V_D = {1 \over 2} g^2 (\phi_1^* \phi_1 - \phi_2^* \phi_2)^2,
\end{equation}
thus enforcing the equality $|v_1| = |v_2|$.  In this case, the scalar 
field $\sqrt 2 Re \zeta$ acquires a nonzero mass from $V_D$, whereas 
$\sqrt 2 Im \zeta$ remains massless.  The latter is of course the familiar 
would-be Nambu-Goldstone boson which gets absorbed by the U(1) gauge boson 
to render the latter massive \cite{6}.  Both it and $\sqrt 2 Re \zeta$ have 
the mass $2 g |v|$, as are their fermionic partners. 

The other is to break the supersymmetry which is certainly necessary 
phenomenologically.  As shown below, this would also fix $v_1$ and $v_2$ 
separately, even if the supersymmetry breaking parameter is very small, 
as long as it is nonzero, {\it i.e.} the scale-invariant supersymmetric 
ground state is inherently \underline {unstable}.

Let $V$ of Eq.~(6) be supplemented with the soft supersymmetry breaking 
term $a |\phi_1|^2$, where $a < 0$.  The minimum of $V$ is now shifted:
\begin{equation}
\langle \chi \rangle = u = -{\mu \over f} + b,
\end{equation}
and
\begin{equation}
\langle \phi_1 \phi_2 \rangle = v_1 v_2 = -{m u \over f} - {h u^2 \over f} 
+ c,
\end{equation}
where
\begin{eqnarray}
v_1 (a + f^2 b^2) + v_2 f^2 c &=& 0, \\ v_2 b^2 + v_1 c &=& 0, \\ 
(v_1^2 + v_2^2) f b + (m + 2 h u)c &=& 0.
\end{eqnarray}
For $a = 0$, it is clear that the only solution is $b = c = 0$.  For 
$a \neq 0$, it becomes
\begin{equation}
b = {v_1 \over f} \sqrt {-a \over v_1^2 - v_2^2}, ~~~ c = {v_1 v_2 a 
\over f^2 (v_1^2 - v_2^2)},
\end{equation}
and
\begin{equation}
{v_1^2 \over v_2^2} - 1 = {-(m + 2 h u)^2 a \over f^4 (v_1^2 + v_2^2)^2}.
\end{equation}
In the above, all parameters have been assumed real for simplicity.  It is 
clear that the mass-scale ambiguity of the supersymmetric theory is now 
resolved.

Whereas the spontaneous supersymmetric generation of an indeterminate mass 
scale and its resolution are not new ideas \cite{Z}, the preceding discussion 
serves to point out an interesting new byproduct.  The fermion mass matrix 
$\cal M$ of Eq.~(16) remains the same in the case of gauging the U(1) 
symmetry, and is approximately the same in the case of softly broken 
supersymmetry.  It has a seesaw texture in that one diagonal entry is zero.  
This is a generic result from the Yukawa coupling of three superfields of 
charges +1, $-1$, and 0.  If the parameters of the theory are such that $|A| 
<< |B|$ in Eq.~(16), one mass eigenvalue is naturally light, {\it i.e.} 
$|A|^2/|B|$, by the well-known seesaw mechanism \cite{7}.  This does require 
some fine tuning, {\it i.e.} $f m \simeq h \mu$.  However, since it involves 
only parameters of the superpotential, the condition is stable against 
radiative corrections.

The appearance of a light Majorana fermion with no standard-model 
interactions may be regarded as a sterile neutrino.  It may be relevant 
to the current experimental situation of neutrino oscillations, where 
positive signals are being claimed in atmospheric \cite{8}, solar \cite{9}, 
and accelerator \cite{10} data.  With three known doublet neutrinos, it 
is difficult to accommodate all three sets of data with sensitivity to three 
very different mass scales.  Hence a fourth singlet neutrino is needed 
\cite{11}.

In conclusion, although it is possible to generate an indeterminate mass 
scale from the spontaneous breakdown of a global continuous symmetry of 
a supersymmetric field theory, the resulting scale-invariant ground state 
is inherently unstable with respect to any soft supersymmetry breaking. 
It also does not happen if the global symmetry is made local.  However, 
an interesting generic feature occurs in the fermion sector, where a 
light sterile neutrino may appear.

\newpage
\begin{center}{ACKNOWLEDGEMENT}
\end{center}

This work was supported in part by the U.~S.~Department of Energy under 
Grant No. DE-FG03-94ER40837.

\bibliographystyle{unsrt}

\end{document}